\documentclass[twocolumn,times,twocolappendix]{aastex631}

\usepackage{amsmath}
\usepackage{multirow}
\usepackage{times}
\usepackage{xcolor}
\usepackage{url}
\usepackage{wrapfig}
\usepackage{graphicx}
\usepackage{bm}
\usepackage{comment}

\received{September 1, 2023}
\revised{December 18, 2023}
\accepted{December 19, 2023}

\submitjournal{ApJ Letters}

\shorttitle{Measuring the tracers of planetary formation in the atmosphere of WASP-77A b}
\shortauthors{Edwards \& Changeat}

\graphicspath{{./}{Figures/}}

\begin{document}

\title{Measuring Tracers of Planet Formation in the Atmosphere of WASP-77A b: Sub-stellar O/H and C/H ratios, with a stellar C/O ratio and a potentially Super-stellar Ti/H ratio}

\correspondingauthor{Billy Edwards}
	\email{b.edwards@sron.nl}
	\author[0000-0002-5494-3237]{Billy Edwards}
	\affil{SRON, Netherlands Institute for Space Research, Niels Bohrweg 4, NL-2333 CA, Leiden, The Netherlands}
 \affil{Department of Physics and Astronomy, University College London, Gower Street, WC1E 6BT London, United Kingdom}
\author[0000-0001-6516-4493]{Quentin Changeat}
\affil{European Space Agency (ESA), ESA Office, Space Telescope Science Institute (STScI), Baltimore MD 21218, USA}
\affil{Department of Physics and Astronomy, University College London, Gower Street, WC1E 6BT London, United Kingdom}
 
\begin{abstract}

We present a comprehensive atmospheric retrieval study of the hot Jupiter WASP-77A\,b using eclipse observations from the Hubble Space Telescope (HST) and JWST. Using atmospheric retrievals, the spectral features of H$_2$O, CO, and TiO are identified, with volume mixing ratios estimated at log$_{\rm 10}$(VMR) = -4.40$^{+0.14}_{-0.11}$, -4.44$^{+0.34}_{-0.28}$, and -6.40$^{+0.22}_{-0.23}$, respectively. We derive the atmospheric carbon-to-oxygen ratio -- a key planetary formation tracer -- to be C/O = 0.54$\pm$0.12, which is consistent with both the stellar host value and previous studies of the planet's atmosphere, suggesting a relatively close-in formation. Computing other elemental ratios (i.e., C/H, O/H, and Ti/H), we conclude that the general enrichment of the atmosphere (i.e., metallicity) is sub-stellar, is depleted in C and O, but that Ti appears slightly super-stellar. A low C and O content could be obtained, in combination with a stellar C/O ratio, if the planet formed outside of the CO$_2$ snow line before migrating inwards. Meanwhile, a super-stellar Ti/H could be obtained by late contamination from refractory rich planetesimals. While broadly in agreement with previous works, we do find some differences and discuss these while also highlighting the need for homogeneous analyses when comparative exoplanetology is conducted.

\end{abstract}

\keywords{Exoplanet atmospheres (487); Hot Jupiters (753);  Hubble Space Telescope (761); JWST (2291)}

\section{Introduction} \label{sec:intro}

Despite being a rare outcome of planetary formation, numerous hot Jupiters have been detected due to the transit technique being biased toward large planets on short orbits. As their size and temperature are favourable for atmospheric characterisation, most atmospheric observational studies using space-based instruments have focused on this class of objects. With the advent of the spatial scanning technique \citep{mccullough_wfc3_scan}, the Wide Field Camera 3 (WFC3) on board the Hubble Space Telescope (HST) has enabled around one hundred of those planets to be characterised via transit \citep[e.g.,][]{tsiaras_30planets, pinhas, cubillos_pop, kawashima_pop, transmission_pop} and eclipse \citep[e.g.,][]{mansfield_metric, emission_pop} spectroscopy, enabling the search for trends in their atmospheric composition. 

More recently, JWST has become the premier facility for space-based exoplanet spectroscopy. The four instruments on board JWST offer a wider simultaneous wavelength coverage than was previously available as well as access to previously uncharted spectral regions. Early studies of giant exoplanets have successfully used each of these JWST instruments for transit \citep[e.g.,][]{dyrek_w107,feinstein_niriss}, eclipse \citep[e.g.,][]{bean_hd149,coulombe_w18} and phase-curve observations \citep[e.g.,][]{bell_w43}. 

Here, we conduct a comprehensive retrieval study on WASP-77A\,b, an inflated hot Jupiter in a wide binary system \citep{maxted_w77}, using data from HST and JWST. The planet orbits WASP-77A, a G8V star. WASP-77 B, a fainter K-dwarf companion to WASP-77A, is separated by 3". WASP-77A\,b has been previously observed in emission with the ground-based high-resolution Immersion GRating INfrared Spectrometer (IGRINS) on Gemini-South (covering wavelengths from $\lambda \in [1.45,2.55]\, \mu$m). Those observations led to tight constraints on the atmospheric metallicity (log$_{\rm 10}$(M/H) = -0.48$^{+0.15}_{-0.13}$) and the carbon-to-oxygen ratio (C/O = 0.59$\pm$0.08) through the measurement of the atmospheric H$_2$O and CO abundances \citep{line_w77}. Hence, the data suggested a metal-poor atmosphere and a solar C/O ratio. 

Low-resolution eclipse observations from the HST Wide Field Camera 3 (WFC3) G141 grism complemented this picture, showing a clear water absorption feature at $\lambda = 1.4\, \mu$m \citep{emission_pop,mansfield_w77} and indicating a dayside positive lapse rate (i.e., a decreasing with altitude thermal structure). The data, however, did not precisely constrain the H$_2$O abundance, nor was it possible to clearly infer the amount of CO in the atmosphere, even when combining it with photometric data from Spitzer \citep{mansfield_w77}. 

As part of the GTO-1274 programme, an eclipse of WASP-77A\,b was captured using the Near-InfraRed Spectrometer (NIRSpec) on JWST. \citet{august_w77_jwst} analysed this data using chemical equilibrium retrievals, concluding that the data was best-fit by a sub-solar metallicity (log$_{\rm 10}$(M/H) = -0.91$^{+0.24}_{-0.16}$) and a low C/O ratio (0.36$^{+0.10}_{-0.09}$) atmospheres. These results roughly agreed with the conclusions from the Gemini data, but the preferred models were not able to fit the HST WFC3 spectrum from \citet{mansfield_w77}. However, the spectrum from \citet{mansfield_w77} is visually at odds with one obtained by \cite{emission_pop}, despite being derived from the same HST data.

In this work we explore the atmospheric properties of WASP-77A\,b, focusing on the recovery of key planetary formation tracers. We use information from the novel eclipse observations by JWST-NIRSpec \citep{august_w77_jwst}, attempting to reconcile the tension between the different HST reductions \citep{emission_pop, mansfield_w77} to provide a comprehensive interpretation of WASP-77A\,b's atmosphere. In Section \ref{sec:methods} we describe the data used in this study and our retrieval setup. We present our results in Section \ref{sec:results} and discuss their implications in Section \ref{sec:discussion}. 

\section{Methodology}
\label{sec:methods}

The emission spectrum of WASP-77A\,b has been captured at low-resolution (i.e., R $<$ 5000) by both HST and JWST. The HST data were taken with the WFC3 G141 grism, giving a spectral coverage of $\lambda \in [1.1,1.6]\, \mu$m. In the main text, we focus on the reduction from \citet{emission_pop} but discuss its robustness in Appendix \ref{sec:appendix_a}, comparing their methodology against that of \citet{mansfield_w77}. The JWST data was acquired with the NIRSpec instrument using the Bright Object Time-Series (BOTS) mode, with the G395H grating and F290LP filter combination. The spectrum analysed in our study is from \citet{august_w77_jwst} and covers $\lambda \in [2.674, 3.716]\, \mu$m (NRS1) and $\lambda \in [3.827, 5.173\, \mu$m (NRS2). More details on the observational setups and data reduction procedure can be found in each of these studies.

We invert the atmospheric properties of WASP-77A\,b from the observed spectra using the publicly available Bayesian retrieval suite \texttt{TauREx 3.1} \citep{al-refaie_taurex3,taurex3_chem}\footnote{\url{https://github.com/ucl-exoplanets/TauREx3_public}}. We assume WASP-77A\,b possesses a primary atmosphere with a solar helium-to-hydrogen ratio (He/H$_2$ = 0.17). The atmosphere is modelled between $p \in [10^{-4}, 10^6]$ Pa using 100 plane-parallel layers uniformly partitioned in log-space. The radiative contributions of the relevant molecules, Collision Induced Absorption (CIA) from H$_2$-H$_2$ \citep{abel_h2-h2, fletcher_h2-h2} and H$_2$-He \citep{abel_h2-he}, and Rayleigh scattering \citep{cox_allen_rayleigh} are included in the model. Stellar and planetary parameters are taken from \citet{maxted_w77}, with the host-star emission being modeled by a PHOENIX spectrum \citep{phoenix}.

We perform two types of retrievals: 1) retrievals where each molecular species is independently fitted for (referred to as {\it free retrievals}), and 2) retrievals that assume a gas mixture at chemical equilibrium via Gibbs free energy minimisation (referred to as {\it equilibrium retrievals}).

1) {\it Free retrievals}: We include the molecular opacities from the ExoMol \citep{Tennyson_exomol,chubb_database}, HITRAN \citep{gordon} and HITEMP \citep{rothman} databases. Considered species are H$_2$O \citep{polyansky_h2o}, CH$_4$ \citep{exomol_ch4}, CO \citep{li_co_2015}, CO$_2$ \citep{rothman_hitremp_2010}, TiO \citep{McKemmish_TiO_new}, VO \citep{mckemmish_vo}, FeH \citep{wende_FeH} and H$^-$. To include H$^-$, we use the description in \citet{edwards_ares} with the coefficients of Table 1 from \citet{john_1988_h-}. For the molecular abundances, uniform priors of log$_{\rm 10}$(VMR) $\in [-15, -1]$ are used.

2) {\it Equilibrium retrievals}: We use the code \texttt{GGchem} \citep{woitke_ggchem} via the \texttt{TauREx 3} plugin system \citep{taurex3_chem} to model the atmospheric chemistry of WASP-77A\,b. The free chemical parameters are: atmospheric metallicity (M/H), the C/O ratio, and the Ti/O ratio. The Ti/O ratio is included following \citet{emission_pop}, who noted an apparent population-wide refractory enrichment for hot Jupiter planets, leading to poor fits of HST WFC3 spectra when assuming solar Ti/O ratio. For those retrievals, the priors are also uniform with log$_{\rm 10}$(M/H) $\in [-2, 2]$, C/O $\in [0.05, 2]$, and log$_{\rm 10}$(Ti/O) $\in [-10, 5]$.

For all our retrievals, we employed a parametric N-point temperature-pressure ($T-p$) profile. Following \citet{changeat_w43}, we retrieve the temperature value of height nodes at fixed pressures ($p \in \{10^1, 10^0, 10^{-1}, 10^{-2}, 10^{-3}, 10^{-4}, 10^{-6}, 10^{-8}\}$ Bar). Such a $T-p$ profile allows the temperature structure of the planet to be determined wholly from the data without any a priori assumptions. 

For HST, recovering absolute transit or eclipse depths is difficult due to its strong instrument systematics \citep[e.g., ][]{guo_hd97658,emission_pop,transmission_pop}. To mitigate for this in retrievals that combine spectral from different instruments, one can fit for an additional mean offset \citep[e.g., ][]{yip_w96} to attempt to alleviate any biases. In our retrievals, we always allow the HST WFC3 G141 spectrum to be shifted relative to the JWST NIRSpec spectrum due to its stronger systematics and the potential residual contamination from WASP-77B when deriving the spectrum \citep{emission_pop, mansfield_w77}. In the high-resolution mode, the NIRSpec data are split across two detectors (NRS1 and NRS2). As there could be offsets between these spectra, we also allow for the NRS2 spectrum to be offset with respect to the NRS1 spectrum\footnote{We also performed retrievals where the NRS1 spectrum was offset and this did not change our results or conclusions.}. For both offsets, the bounds for the offset ($\Delta$) are set to be extremely broad with $\Delta \in [-500, +500]$ ppm.

We explore the parameter space using the nested sampling algorithm \texttt{MultiNest} \citep{Feroz_multinest,buchner_multinest} with 1000 live points and an evidence tolerance of 0.5. 

The free chemical retrievals serve to derive elemental ratios for WASP-77A\,b, which are then compared with the host-star values. With respect to solar values, \citet{polanski_st_abund} measured C/H = -0.02, O/H = 0.06, and Ti/H = 0.01. Using the solar abundance for these elements \citep{asplund_solar_abund}, these yield C/O = 0.46 and Ti/O = 1.6$\times$10$^{-4}$. We compare those values to the derived/fitted ratios from our retrievals. In Section \ref{sec:diff_studies}, we discuss the implication of comparing to the stellar abundances derived in other studies.

\section{Results}
\label{sec:results}

The joint WFC3+NIRSpec fit shows evidence for three molecules: H$_2$O, CO, and TiO. The preferred abundances for these species are log$_{\rm 10}$(H$_2$O) = -4.58$^{+0.16}_{-0.13}$, log$_{\rm 10}$(CO) = -4.51$^{+0.28}_{-0.26}$, and log$_{\rm 10}$(TiO) = -6.52$^{+0.22}_{-0.23}$. The molecular features are seen in absorption, indicating a positive lapse rate (see Figure \ref{fig:spectrum_and_tp}) with no evidence for a stratosphere. The free retrieval prefers the application of a small offset {($\Delta$ = -39$^{+22}_{-20}$\,ppm and -56$^{+36}_{-33}$\,ppm) to the WFC3 and NRS2 spectra but, as shown by the posterior distributions in Figure \ref{fig:posteriors}, these parameters do not have a strong correlation with the molecular abundances.

Using the retrieved abundances of H$_2$O, CO, and TiO, we compute the following ratios to 1$\sigma$: C/O = 0.54$\pm$0.12 and log$_{\rm 10}$(Ti/O) = -2.30$^{+0.20}_{-0.23}$. Therefore, the planetary C/O ratio is compatible both with that of WASP-77A \citep[C/O = 0.46, ][]{polanski_st_abund}, and the solar value \citep[0.55,][]{asplund_solar_abund}. However, we find the Ti/O ratio to be much higher for the planet than for the host star \citep[log$_{\rm 10}$(Ti/O) = -3.79, ][]{polanski_st_abund}.

We also determine the elemental ratios with respect to hydrogen. For C/H and Ti/H, only the CO abundance and the TiO abundance are used, respectively. For O/H, all three molecules are used. These ratios were normalised to the solar values from \citet{asplund_solar_abund} and the stellar values from \citet{polanski_st_abund}. 

\begin{figure*}
     \centering
     \includegraphics[width=0.975\textwidth]{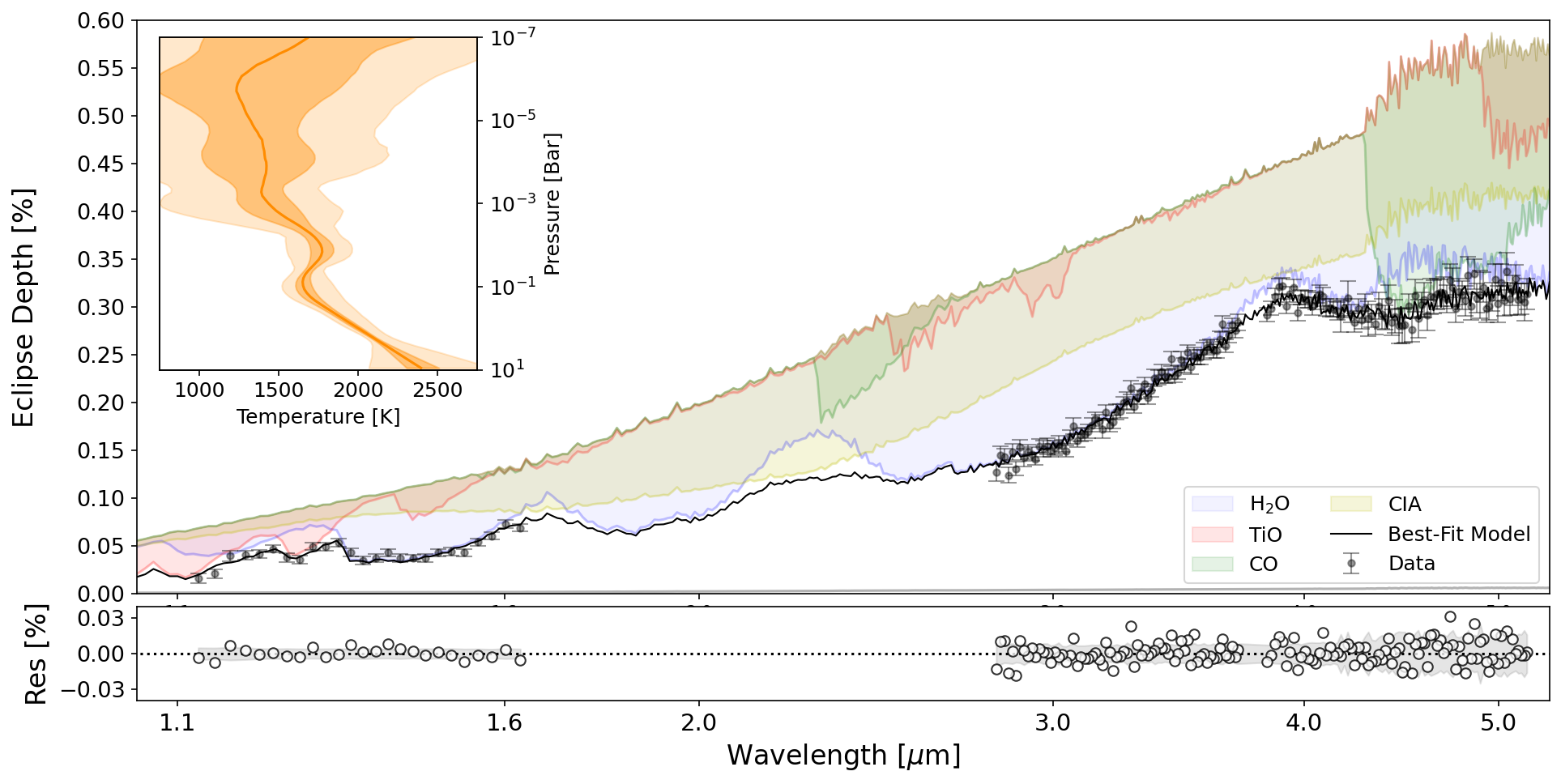}
     \caption{Best-fit free retrieval of the JWST NIRSpec and HST WFC3 data for WASP-77A\,b. The coloured regions show the individual contribution of three species (H$_2$O, CO, and TiO) and Collision Induced Absorption (CIA), while the inset shows the 1, 2, and 3$\sigma$ confidence intervals for the retrieved temperature-pressure ($T-p$) profile. These three species are detected in the HST and JWST data, allowing us to place constraints on elemental ratios for this atmosphere. The best-fit $T-p$ profile has a positive lapse rate (i.e, no evidence for a thermal inversion) in the pressure region probed by our observations ($p \in [10^5, 100]$ Pa).}
     \label{fig:spectrum_and_tp}
\end{figure*}

\begin{figure*}
     \centering
     \includegraphics[width=0.75\textwidth]{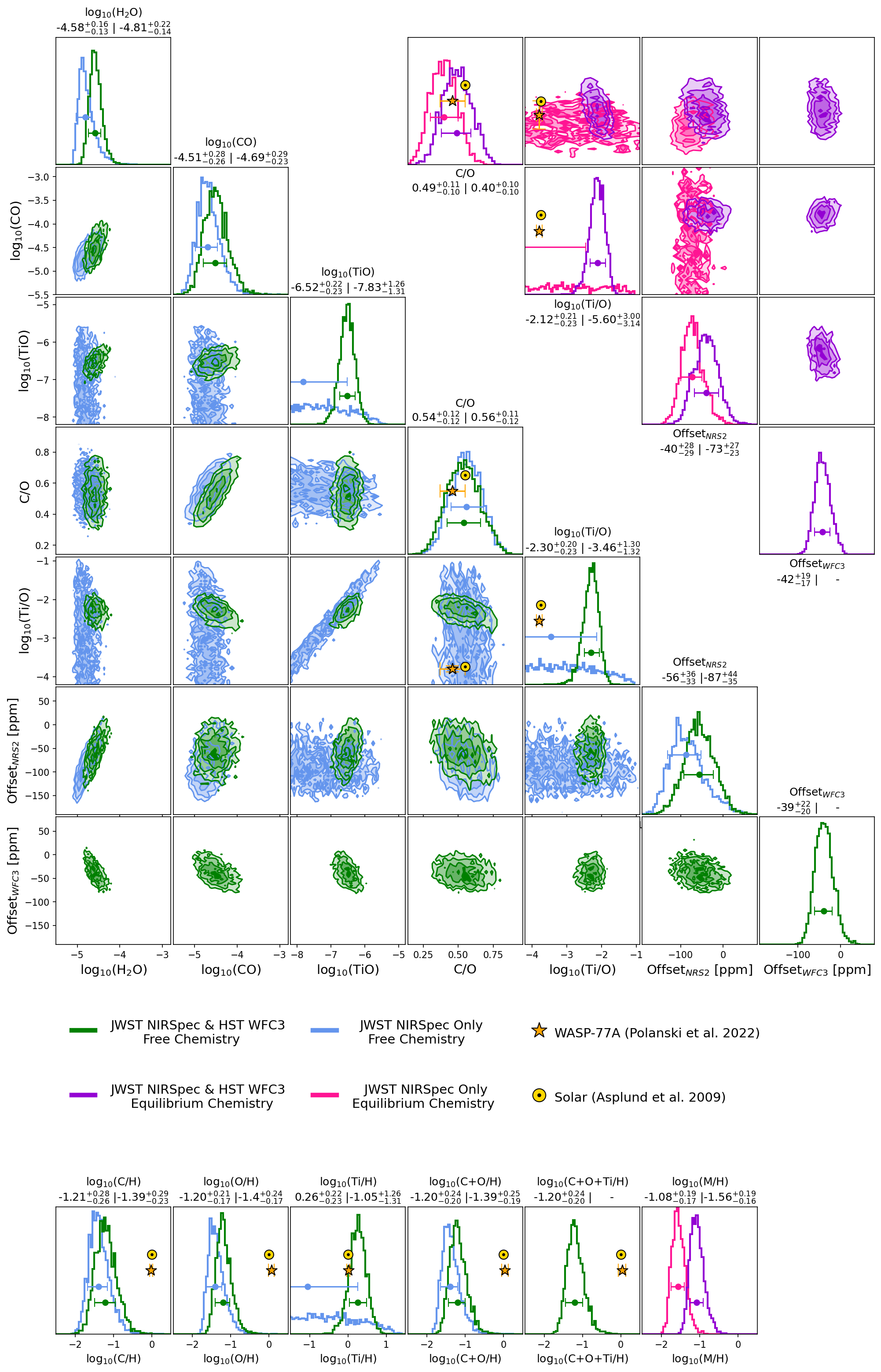}
     \caption{Posterior distributions for the dayside of WASP-77A\,b. For the free chemistry retrievals, the C/O and Ti/O ratios are derived parameters. Where there are two reported values, they are given as those from the combined JWST NIRSpec and HST WFC3 fit (left) and the JWST NIRSpec only (right). Both free and equilibrium chemistry models, as well as fits to both datasets or just JWST NIRSpec, prefer a roughly stellar C/O ratio. The free and equilibrium retrievals to the combined datasets also prefer a distinctly super-stellar Ti/O ratio. All elemental-to-hydrogen abundances are given with respect to solar values. Again, C/H and O/H are sub-stellar whereas Ti/H is super-stellar. All ratios and abundances are computed in terms of the volume mixing ratio.}
     \label{fig:posteriors}
 \end{figure*}

We find that C/H and O/H are both clearly sub-stellar, with C/H = 0.06$^{+0.06}_{-0.03}\,\times$stellar and O/H = 0.06$^{+0.04}_{-0.02}\,\times$stellar. However, we find a slightly super-stellar Ti/H ratio: Ti/H = 1.77$^{+1.15}_{-0.72}\,\times$stellar. The apparent enrichment in titanium compared to the stellar value is far smaller when considering Ti/H than Ti/O, due to the simultaneous sub-stellar prevalence of oxygen. All these ratios are given in Table \ref{tab:ratios}, which can be found in Appendix \ref{sec:appendix_b}.

As our retrievals indicate a depletion in carbon and oxygen for WASP-77A\,b's atmosphere, the metallicity when calculated from these elements is also sub-stellar with (C+O)/H = 0.06$^{+0.04}_{-0.02}\,\times$stellar. However, this contrasts with the retrieved abundance of Ti that is suggestive of a super-stellar Ti/H ratio. Nevertheless, accounting for Ti in the metallicity by using (C+O+Ti)/H makes little difference due to the low abundance of Ti compared to the other elements (see Table \ref{tab:ratios}).

While the free retrieval to both datasets is our preferred approach due to its unassuming nature, we use additional retrievals to explore the robustness of our results. 

Firstly, we conduct an equilibrium retrieval on the joint dataset, finding similar, though slightly lower, C/O and Ti/O ratios to the free retrieval: C/O = 0.49$^{+0.11}_{-0.10}$ and log$_{\rm 10}$(Ti/O) = -2.12$^{+0.21}_{-0.23}$. Comparing the metallicity (M/H) again shows excellent agreement, with log$_{\rm 10}$(M/H) = -1.08$^{+0.19}_{-0.17}$ (M/H = 0.09$\pm$0.03$\,\times$stellar). Hence, free and equilibrium retrievals on the WFC3+NIRSpec data show a consistent picture, strongly suggesting that WASP-77A\,b hosts a low metallicity atmosphere, a conclusion which is in line with previous studies. They also both suggest a slight enrichment of Ti.

Secondly, both free and equilibrium retrievals are performed on the NIRSpec spectrum alone. As the spectral features of CO and H$_2$O are prominent in the NIRSpec spectrum, the free retrieval infers similar abundances for those molecules to the combined fit. Hence, the derived C/O ratio remains similar, suggesting that the NIRSpec data drive our conclusions for these species. The equilibrium retrieval finds a similar metallicity but a lower C/O ratio. However, as TiO does not have broadband spectral features in the NIRSpec G395 spectral range, both free and equilibrium retrievals could not place constraints on TiO. Therefore, the Ti/O ratio remains unconstrained in the NIRSpec-only retrievals.

Comparisons of all these retrievals are shown in Figure \ref{fig:posteriors}, including the derived elemental ratios. Each model led to a C/O ratio that was consistent to 1$\sigma$ with the stellar value \citep{polanski_st_abund} and to a definitively sub-stellar metallicity. The retrievals that included both NIRSpec and WFC3 data were consistent with an atmosphere enriched in titanium when compared to the host star.

\section{Discussion}
 \label{sec:discussion}

\subsection{Potential implications for planet formation}
\label{sec:planet_form}

Elemental ratios -- such as those derived in this study -- have long been proposed as potential tracers of planetary formation and evolution. The most widely considered tracers are the bulk metallicity (i.e., M/H) and the C/O ratio \citep[e.g.,][]{oberg_co, mordasini_2016, madhu_formation, Eistrup_2018, Cridland_2019}, although more recently, other tracers -- such as S/O, N/O \citep[e.g.,][]{turrini_formation,pacetti_formation, Ohno_2023a, Ohno_2023b}, or refractory-to-O \citep[labelled R/O:][]{lothringer_ref_ratio} -- have also been suggested to break the degeneracy in current formation models. 

Suggestions of super-stellar metallicities for close-in giant planets from observational studies \citep[e.g.,][]{thorngren_2016,fortney_ch4,bean_hd149,feinstein_niriss} have motivated a growing body of works to explain their formation by forming far out in the protoplanetary disk before undergoing extensive disk migration, which is coupled with the efficient accretion of planetesimals and gas-enriched materials at the disk snow-lines \citep{Booth_2017, Hasegawa_2018, Shibata_2020, turrini_formation, pacetti_formation, Khorshid_2022, Schneider_2021}. However, this picture does not fully explain the diversity of hot Jupiter compositions, with evidence for sub-stellar O/H exoplanets and trends for high refractory content also being found \citep{emission_pop}. Clearly, WASP-77A\,b does not appear to be enriched in volatiles (O and C) relative to its host star but could possess a high R/O ratio (here probed by Ti/O), thus suggesting an alternative pathway to its formation.  

The roughly stellar C/O suggested by our retrievals implies that WASP-77A\,b might have formed around the H$_2$O/CO$_2$ ice lines (i.e, relatively close-in formation), as one would expect an enriched C/O ratio \citep[$>$0.8, ][]{oberg_co,madhu_formation} if significant accretion had occurred beyond the snow lines. However, for these lower C/O ratios, models also usually predict super-stellar C and O abundances \citep[e.g.][]{Schneider_2021}, which we do not find here: our estimate of the C/H ratio, for instance, is far below the stellar ratio as well as that of the Solar System gas giants \citep{atreya_ss_met}. A low C and O content could be obtained if the planet formed mainly from gas (e.g., from gravitational instabilities rather than core accretion), but this is believed to only occur far out in the protoplanetary disk. One formation pathway that could explain a low C and O content, combined with a stellar C/O ratio, is if WASP-77A\,b planet formed exterior to the CO$_2$ evaporation front and only crossed the CO$_2$ snow line, which is located at around 15 AU, very late in its evolution \citep{Bitsh_2022}.

A high R/O ratio (e.g., here Ti/O) also implies a close-in formation rather than extensive migration after the disk dispersal and could indicate complex interactions at the snow lines involving evaporating pebbles \citep{Schneider_2021, Bitsh_2022}. As our retrievals suggest a low prevalence of oxygen and carbon with a potentially super-stellar abundance of titanium (i.e., a high R/H ratio), WASP-77A\,b could be well explained by a formation around the CO$_2$ ice line with late enrichment of its atmosphere by rock-rich planetesimals \citep{lothringer_ref_ratio,Bitsh_2022}. Data that allow us to constrain other R/O ratios (e.g., Si/O via SiO) may yield further insights into the formation pathway for this planet by independently measuring the enrichment of those elements as well as adding sensitivity to additional oxygen reservoirs, reducing the chance of an oxygen deficit in our calculations \citep{fonte_oxygen}.

\subsection{Strength of the TiO Detection}

Evidence for TiO has been found in the HST WFC3 emission spectra of several other planets \citep[e.g.,][]{Haynes_Wasp33b_spectrum_em,edwards_ares,changeat_edwards_k9}. However, these claims have sometimes been disputed by independent analyses \citep[e.g.,][]{jacobs_k9} or by non-detections via high-resolution spectroscopy \citep[e.g.,][]{merritt_w121,kasper_k9}. For WASP-77A\,b, the high-resolution study by \citet{line_w77} did not consider TiO, likely due to the lack of sensitivity of their data to this molecule due to the observed wavelengths ($\lambda \in [1.43, 2.42]\, \mu$m).

Often, the ``detection'' of optical absorbers with WFC3 G141 stems from only one or two data points. However, for WASP-77A\,b, the detection is supported by two features: one below $\lambda = 1.15\, \mu$m (i.e., at the edge of the WFC3 G141 bandpass) and a second at $\lambda \in [1.23,1.28]\, \mu$m (see Figure \ref{fig:tio_comp}). Performing an equilibrium retrieval which assumes a solar Ti/O ratio smooths the TiO absorption features in those regions and leads to a poorer fit of the WFC3 data, with $\Delta$ln(E) = 6.8 in favour of the model with super-solar Ti/O (i.e., a 4.2$\sigma$ preference). Note that, since the data strongly favours a sub-stellar O/H ratio that is driven by the more precise NIRSpec data, a solar Ti/O ratio enforces a distinctly sub-stellar Ti/H ratio, when our preferred enhanced Ti model is only slightly super-solar. Therefore, this suggests Ti/H may be a more relevant marker of planetary formation processes than Ti/O.

While the Bayesian evidence points toward the presence of TiO, the combination of this molecule and a lack of a stratosphere is unexpected for highly-irradiated atmospheres. TiO, and other oxides and hydrides, are strong absorbers of visible light and are therefore expected to cause thermal inversion \citep[e.g.,][]{hubeny_tio} by the deposition of stellar energy at high altitudes. Cold traps, where the temperature profile dips below the condensation curve of a molecule, have been suggested as a way of sequestering TiO from the atmosphere \citep[e.g.,][]{spiegel_tio,beatty_k13}. Usually, these models predict the cold-trap to lie between a deeper, warmer atmospheric layer where TiO is present and the stratosphere where again temperatures are hot enough for TiO to avoid condensation. While we do not detect evidence for a thermal inversion, the large uncertainty in the $T-p$ profile at high altitudes (i.e., p $<$ 100 Pa) does not strongly rule out the presence of one either.

\begin{figure}
    \centering
    \includegraphics[width=\columnwidth]{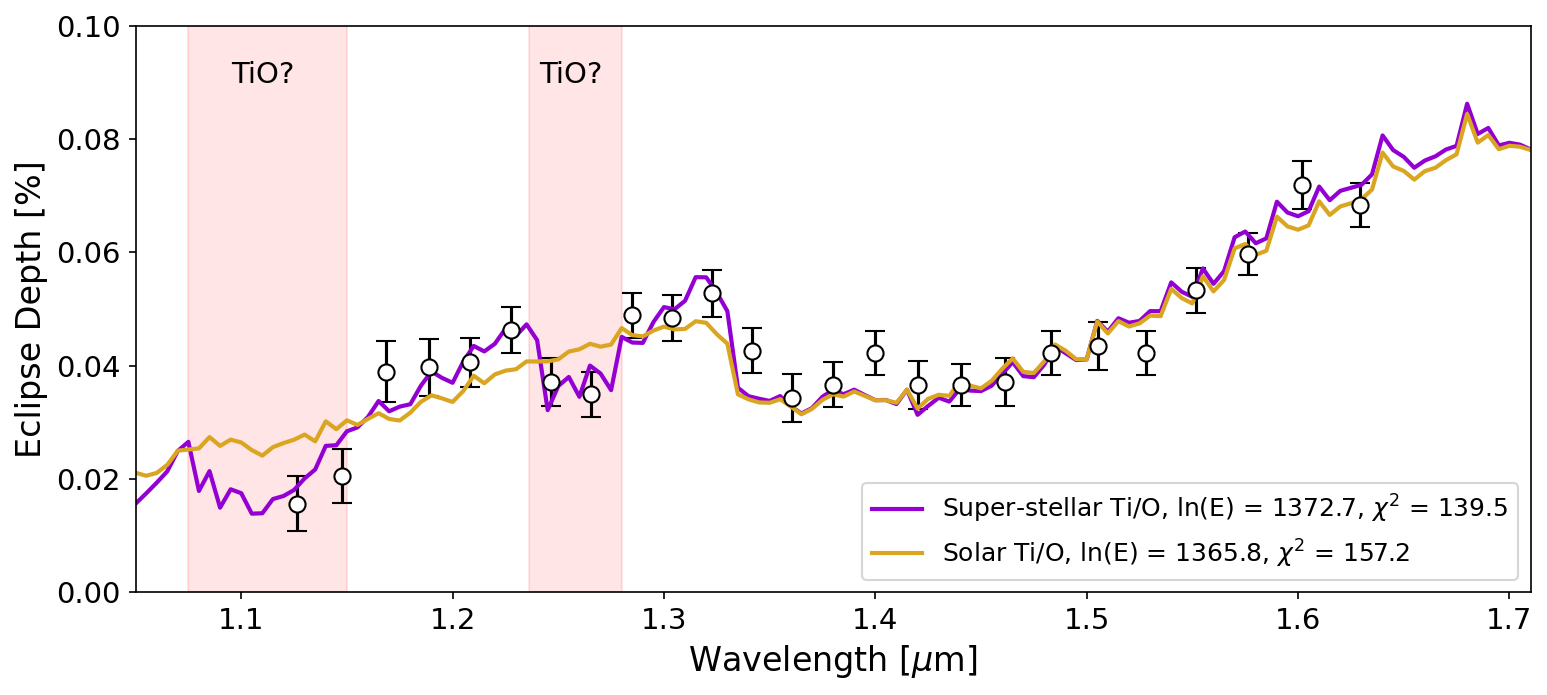}
    \caption{Comparison of the chemical equilibrium retrieval results when we allow the Ti/O ratio to vary (purple) and when we fix it to solar (yellow). The latter does not provide a good fit to the two highlighted regions, and our retrievals suggest the features in these regions are due to TiO.}
    \label{fig:tio_comp}
\end{figure}

\subsection{Comparisons to other works and implications for comparative planetology}\label{sec:diff_studies}


Multiple studies have made independent inferences for the dayside chemistry and thermal structure of WASP-77A\,b using several instruments. Prior studies of the high-resolution Gemini observations \citep{line_w77} and the JWST NIRSpec data \citep{august_w77_jwst} both suggested an atmosphere depleted in carbon and oxygen. However, they slightly disagreed on the value of the C/O ratio. These studies, as well as those of the HST WFC3 spectrum \citep{emission_pop, mansfield_w77}, find a $T-p$ profile with a positive lapse rate. Our joint fit to the HST and JWST data leads to the same general conclusions though with some differences that are worth discussing.

In Figure \ref{fig:ratio_comp}, we compare the molecular abundances and elemental ratios inferred from the WFC3+NIRSpec joint free fit to those from these previous works. Our retrieved abundances of H$_2$O and CO are significantly lower than those reported in \citet{line_w77}. Hence, while both works support the conclusion that the atmosphere of WASP-77A\,b is depleted in C and O, the level of depletion is somewhat different.

\begin{figure}
     \centering
     \includegraphics[width=0.95\columnwidth]{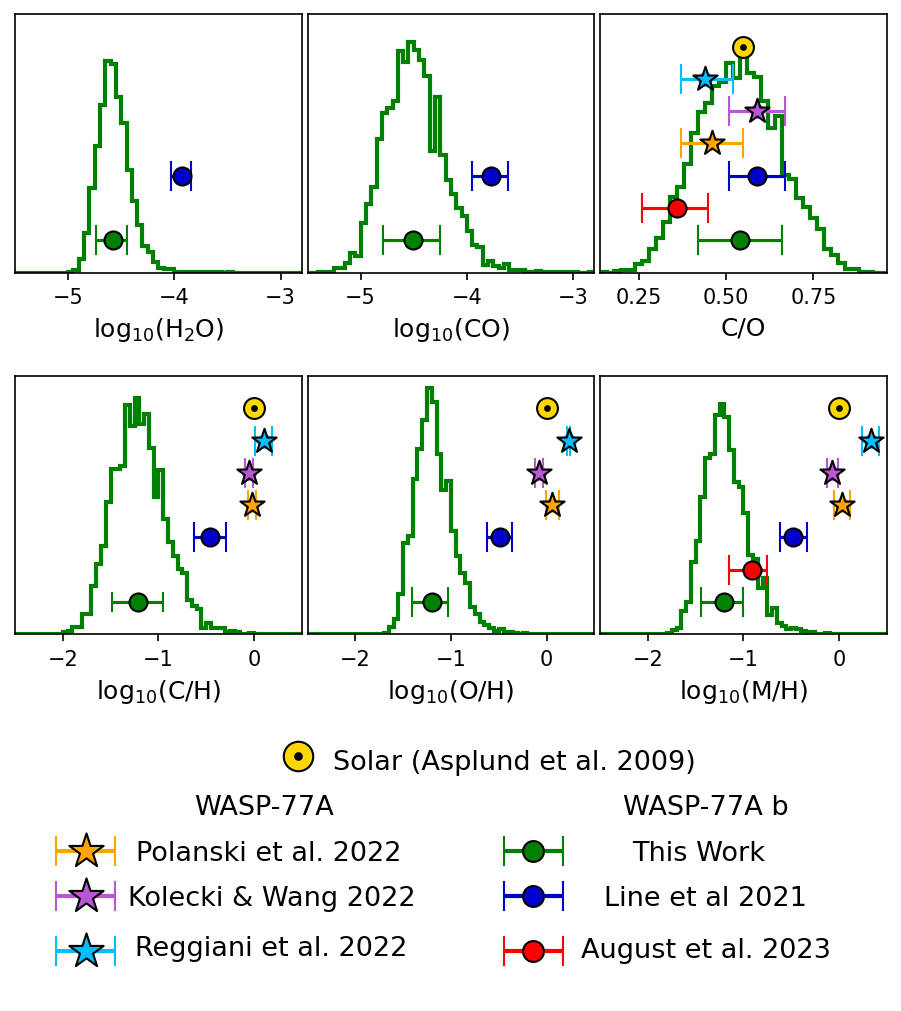}
     \caption{Comparison between the molecular abundances and elemental ratios found here for WASP-77A\,b and those found by literature works. No work has previously tried to constrain the Ti/H ratio, so it is not plotted here. We note that M/H refers to (C+O)/H for \citet{line_w77}, M/H for \citet{august_w77_jwst}, and (C+O+Ti)/H for the atmospheric constraints from this work as well as the measurements of the enrichment of the host star, WASP-77A. }
     \label{fig:ratio_comp}
\end{figure}

Additionally, the preferred C/O ratio from our retrievals lies between those derived by \citet{line_w77} and \citet{august_w77_jwst}, agreeing with both to within 1$\sigma$. The difference with \citet{august_w77_jwst} comes from the addition of the WFC3 data. When performing the chemical equilibrium retrieval on only the NIRSpec data, we also obtain a lower C/O ratio (0.40$\pm$0.1) though it is still within 1$\sigma$ of the stellar value from \citet{polanski_st_abund}. \citet{august_w77_jwst} noted that a reasonable joint fit with the WFC3 data from \citet{mansfield_w77} could not be achieved. As we find no such issue with the spectrum from \citet{emission_pop}, we explore the cause of this in Appendix \ref{sec:appendix_a}.

Furthermore, it is worth noting that \citet{line_w77} analysed the IGRINS data with two different retrieval codes: CHIMERA \citep{line_chimera} and HyDRA-H \citep{hydra_h}. The C/O ratio using their CO and H$_2$O abundances (ED Fig. 6) from these fits are 0.59 and 0.38, respectively. Hence, while their main result claims that WASP-77A\,b has a solar C/O ratio, their secondary result finds a sub-solar C/O ratio that agrees with the JWST NIRSpec-only analyses from this work and \citet{august_w77_jwst}. 

Using the main results of \citet{line_w77}, studies have implied that WASP-77A\,b must have formed far out in the disk beyond the H$_2$O ice line \citep{reggiani_w77,khorshid_w77}. Such a formation was inferred based upon sub-stellar carbon and oxygen abundances and a super-stellar C/O ratio, a condition which \citet{oberg_co} suggested as a unique signature of this formation process. However, in our work, and in that of \citet{august_w77_jwst}, the C/O is not found to be definitively super-stellar and is instead consistent with, or below, the stellar value. If such findings are confirmed (i.e, stellar C/O but depletion in C and O), different planetary formation and evolution mechanisms are required to explain the existence of WASP-77A\,b.


Furthermore, disagreement can occur depending upon the study from which the stellar abundances are taken. Several works have analysed the host star \citep{kolecki_w77, polanski_st_abund, reggiani_w77} and their conclusions also do not always agree. As shown in Figure \ref{fig:ratio_comp}, the C/O ratios for WASP-77A differ between studies,\footnote{We note that \citet{reggiani_w77} suggested that the higher C/O ratio derived in \citet{kolecki_w77} was due to the latter study neglecting extinction in their analysis and that the values from \citet{polanski_st_abund} and \citet{reggiani_w77} are in good agreement for this elemental ratio.}, as do the C/H and O/H ratios. Hence, the inferences made about the planet's formation will clearly change depending upon the study from which the stellar proprieties are derived as well as the atmospheric study of WASP-77A\,b. As such, it becomes extremely difficult to {\it exactly} pinpoint the formation scenario for WASP-77A\,b. These differences, and thus issues, are highlighted in Figure \ref{fig:co_ratio_variation} which shows the deviation from the stellar C/O ratio depending on different datasets and analyses for WASP-77A and WASP-77A\,b.

Clearly, studying a single planet alone will not be sufficient to place strict constraints on the dominant formation and evolutionary pathway for hot Jupiters. Instead, by observing many objects, correlations can be sought between their bulk parameters and the atmospheric chemistry, thereby perhaps shedding light on the processes that dominate their formation at a population level. Such comparative planetology has been a long-held desire within the field of exoplanetary atmospheres \citep[e.g.,][]{cowan_agol,tessenyi}. Population studies of exoplanet atmospheres have been conducted with HST and Spitzer \citep[e.g.,][]{pinhas,baxter_spitzer,kawashima_pop,emission_pop,transmission_pop}, comparisons are already being made between the inferences of JWST data for different planets \citep[e.g.,][]{august_w77_jwst,bean_hd149}, and future missions are being constructed specifically for this task \citep[e.g.,][]{tinetti_ariel,TwinkleSPIE}.

\begin{figure}
    \centering
    \includegraphics[width=0.95\columnwidth]{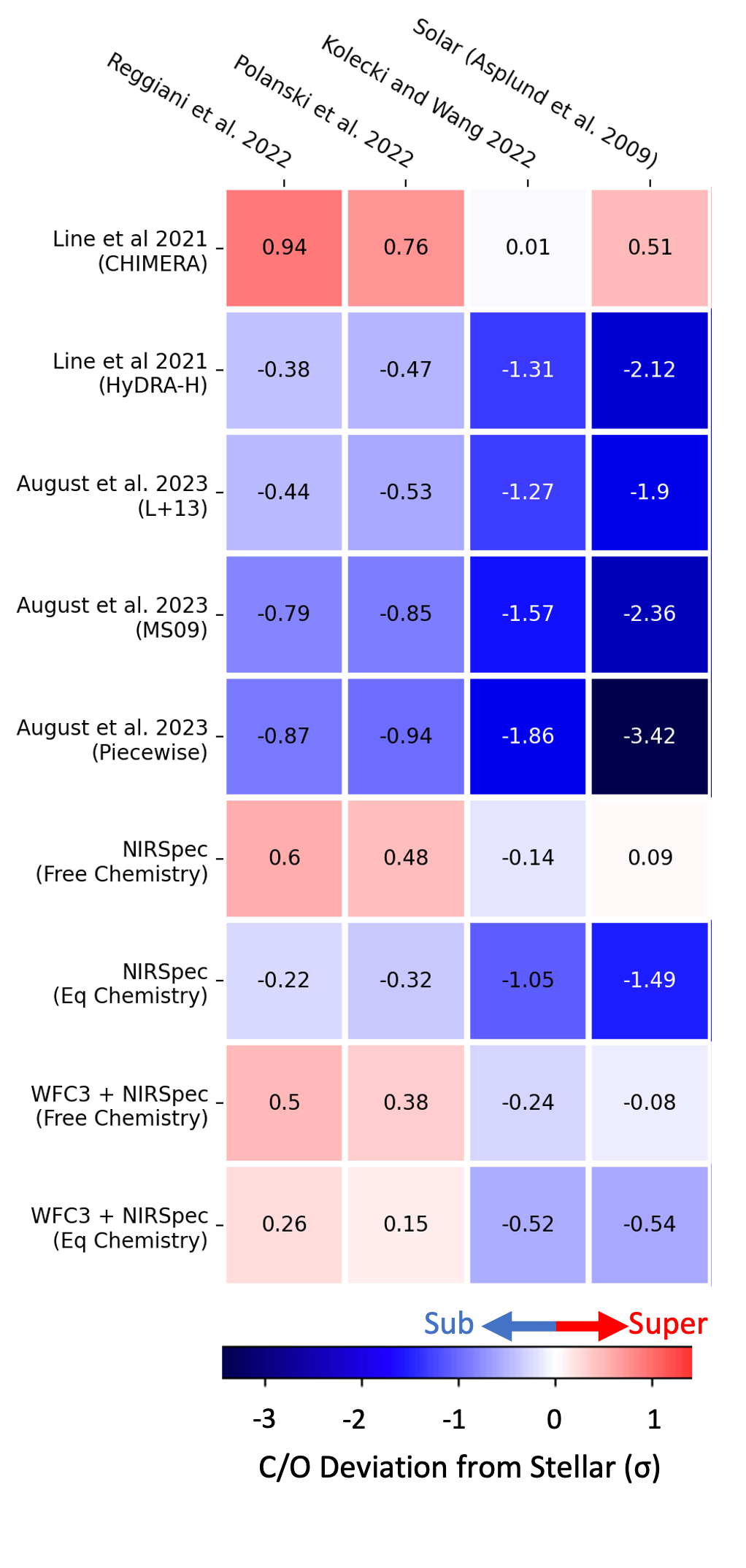}
    \caption{The deviation from stellar C/O ratio for each atmospheric study of WASP-77 A b when considering different stellar abundances. While it is no longer applicable in this case, we also show the comparison to solar abundances as these are the reference values if no stellar measurements have been made. Most studies prefer a sub-stellar (blue squares) or stellar (whiter squares) C/O ratio. The exception is the CHIMERA retrievals from \citet{line_w77} although we note that they are all consistent with all of the stellar values to 1$\sigma$ due to the large uncertainties on the planetary C/O ratio. }
    \label{fig:co_ratio_variation}
\end{figure}

Thus far, most population studies have utilised heterogeneous datasets, comparing the implied chemistry from a variety of studies and instruments. However, without homogeneity, one may invoke a correlation that exists purely because of the sensitivity of the datasets used: it is clear from this study, as well as those from the literature \citep[e.g.,][]{pinhas, aresIII}, that different datasets can lead to different atmospheric compositions for the same planet. While it is clear that homogeneity will be the key to comparative planetology, we note that homogeneity is by no means a guarantee of success: biases will no doubt still be present, and the data will never be sensitive to all species.



\section{Conclusions}


A joint fit to JWST NIRSpec and HST WFC3 spectra of WASP-77A\,b's atmosphere suggests a depletion in carbon and oxygen with respect to stellar values, albeit with a C/O ratio that remains consistent with the stellar value as well as evidence for a slightly super-stellar Ti/H ratio. From a formation perspective, these results suggest that WASP-77A\,b may have formed outside of the CO$_2$ snow line, before migrating inwards and accreting significant rock-rich planetesimals in its atmosphere. Further constraining Ti -- as well as other refractory species: Fe, V, Al, etc. -- will allow us to more confidently assess the claim for super-solar refractory abundance in WASP-77A\,b. In this new era, the unprecedented constraints on elemental ratios offered by JWST data should help us narrow down the diversity of formation pathways for hot Jupiter exoplanets.

\section*{Acknowledgements}

The authors wish to thank Dr. Kazumasa Ohno and Dr. Tadahiro Kimura for their insightful comments and discussion on the potential formation pathways for WASP-77A\,b.\vspace{2mm}

\large{\textit{Funding:}} \normalsize{ Q.C. is funded by the European Space Agency (ESA) under the 2022 ESA Research Fellowship Program. B.E. received travel support for this work under the ESA Science Faculty Funds, funding reference ESA-SCI-SC-LE117. The project also received funding from the Science and Technology Funding Council (STFC) grants ST/S002634/1 and ST/T001836/1.}\vspace{2mm}

\large{\textit{Computing:}} \normalsize{The FI, DIRAC, and OzSTAR national facility at Swinburne University of Technology provided the computing resources; this work utilized the Cambridge Service for Data-Driven Discovery (CSD3), part of which is operated by the University of Cambridge Research Computing on behalf of the STFC DiRAC HPC Facility (www.dirac.ac.uk). The DiRAC component of CSD3 was funded by BEIS capital funding via STFC capital grants ST/P002307/1 and ST/R002452/1 and STFC operations grant ST/R00689X/1. DiRAC is part of the National e-Infrastructure. The OzSTAR program receives funding in part from the Astronomy National Collaborative Research Infrastructure Strategy (NCRIS) allocation provided by the Australian Government.} \vspace{2mm}

\large{\textit{Data:}} \normalsize{This work is based upon publicly available observations taken with the NASA/ESA Hubble Space Telescope \citep[GO-16168, PI: Megan Mansfield; ][]{mansfield_proposal} and the NASA/ESA/CSA JWST \citep[GTO-1274, PI: Jonathan Lunine; ][]{lunine_jwst_prop}. These observations were facilitated by the Space Telescope Science Institute, which is operated by the Association of Universities for Research in Astronomy, Inc., under NASA contract NAS 5–26555. The HST and JWST data presented in this paper were obtained from the Mikulski Archive for Space Telescopes (MAST) at the Space Telescope Science Institute and can be accessed via DOI: \url{10.17909/bsnh-fz53}. We took the NIRSpec spectrum from Table 1 of \citet{august_w77_jwst} and the WFC3 spectrum from \citet{emission_pop}\footnote{\url{https://github.com/QuentChangeat/HST_WFC3_Population/tree/main/Eclipse/Spectra/Eclipses}}.

\facilities{Hubble Space Telescope (WFC3), JWST (NIRSpec).}

\software{TauREx3 \citep{al-refaie_taurex3}, TauREx GGChem \citep{woitke_ggchem,taurex3_chem}, Multinest \citep{Feroz_multinest,buchner_multinest},  Iraclis \citep{tsiaras_hd209}, PyLightcurve \citep{tsiaras_plc}, Astropy \citep{astropy:2013, astropy:2018, astropy:2022}, h5py \citep{hdf5_collette}, emcee \citep{emcee}, Matplotlib \citep{Hunter_matplotlib},  Pandas \citep{mckinney_pandas}, Numpy \citep{oliphant_numpy}, SciPy \citep{scipy}, corner \citep{corner}.}

\bibliography{main}{}
\bibliographystyle{aasjournal}






\appendix

\section{Contamination Correction for the HST WFC3 Data of WASP-77A\,b}
\label{sec:appendix_a}

Here we discuss the reduction of the HST WFC3 spatial scanning data of the eclipse of WASP-77A\,b and the complexities caused by WASP-77B, a fainter \citep[by $\sim$2 magnitudes; ][]{maxted_w77} K-dwarf companion to WASP-77A with a 3" separation. On the detector, the spectra from WASP-77A and WASP-77B overlap, which would adversely affect the recovered emission spectrum for the planet if the contamination is not corrected for.\footnote{Note that the JWST NIRSpec data are not affected by similar issues because the slit (1.6" x 1.6") and the diffraction-limited point spread function (0.17") are both smaller than the separation between the two stars \citep[$\sim$3",][]{maxted_w77,august_w77_jwst}.} Two HST visits were taken under programme GO-16168 \citep[PI: Megan Mansfield; ][]{mansfield_proposal}, each covering an eclipse. The separation of the stars on the detector was different in each case. Two independent studies using a different methodology to handle the contamination from WASP-77B have previously reduced the emission spectrum of WASP-77A\,b: \citet{emission_pop} and \citet{mansfield_w77}.

\citet{emission_pop} used Wayne, a specialised WFC3 simulator \citep{varley_wayne,tsiaras_wfc3_systematics}, to model the contribution of the secondary star. First, they extracted the high-resolution spectra of each star using Iraclis \citep{tsiaras_hd209}. These spectra, and the observational setup of the real data, were then used as inputs to Wayne, which simulated high-resolution, spatially scanned WFC3 images for each star individually. By determining the flux from each star falling in each spectral bin, they determined the flux ratio between the two stars and applied this as a correction to the eclipse spectrum of WASP-77A\,b for each visit.

Meanwhile, \citet{mansfield_w77} utilised a different methodology. At the beginning of each visit, a 0.556\,s staring mode spectrum was taken with the G141 grism. In this data, the spectra from the stars were not overlapping, so the flux of each star could be independently extracted. They then used these spectra to determine the fluxes of the WASP-77A and WASP-77B in each bandpass and corrected the extracted flux using these. Next, they fitted the corrected light curves to obtain the emission spectrum. In their work, \citet{mansfield_w77} noted that their atmospheric models struggled to explain the spectrum and attempted fits in which they deleted several of the data points in an attempt to improve the fit.

\begin{figure*}
    \centering
    \includegraphics[width=0.95\textwidth]{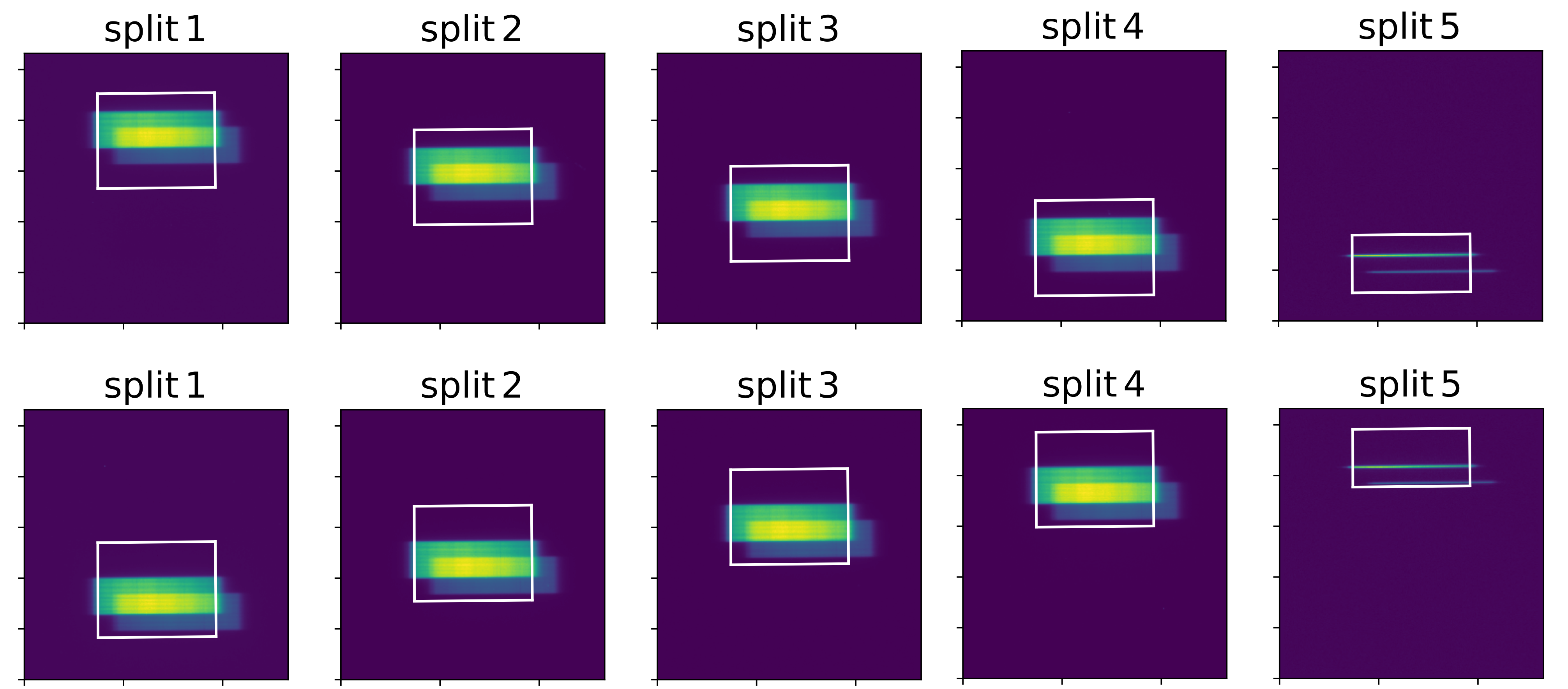}
    \caption{Example extraction apertures when splitting the WFC3 data by the non-destructive reads from the reverse (top) and forward (bottom) scans. The extraction is clearly still contaminated by the fainter secondary star. When the light curve from this extraction is fitted, the spectrum (see Figure \ref{fig:mansfield_comp}) closely matches that from \citet{mansfield_w77}, suggesting that their spectrum is still suffering from contamination.}
    \label{fig:contaminated_extraction}
\end{figure*}

\begin{figure*}
    \centering
    \includegraphics[width=0.95\textwidth]{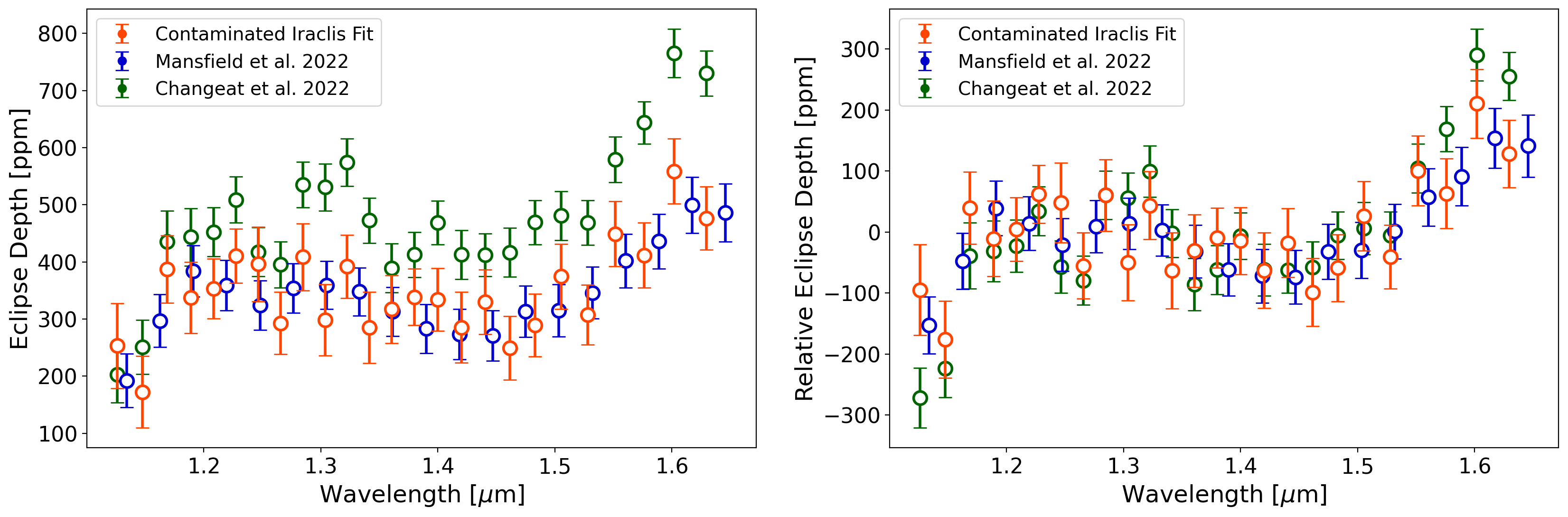}
    \caption{Comparison between different HST WFC3 G141 reductions. The contaminated fit from the extraction shown in Figure \ref{fig:contaminated_extraction} matches the spectrum from \citet{mansfield_w77}. The spectrum from \citet{emission_pop}, which used Wayne to model the contamination, has similar features but a different slope, being significantly deeper at longer wavelengths. \textbf{Left:} absolute eclipse depth. \textbf{Right:} relative eclipse depth having subtracted the mean depth.}
    \label{fig:mansfield_comp}
\end{figure*}

In their work, \citet{august_w77_jwst} noted a disagreement between the HST WFC3 spectrum from \citet{mansfield_w77} and their atmospheric models derived from the JWST NIRSpec data only. When attempting a joint fit, they found their models poorly fitted the HST data. The issue seemed to stem from the spectral shape of the WFC3 data rather than the absolute eclipse depth. An absolute offset is commonly found when using different models for the systematics seen in HST data, but the spectral shape is generally conserved \citep{aresV,transmission_pop}. Therefore, they suggested that perhaps the HST WFC3 data could not be trusted but did not investigate the causes of this discrepancy. 

Here, we compare the spectra from \citet{mansfield_w77} and \citet{emission_pop}, finding differences in their spectral shape. Given the robustness of the spectral shape to pipeline assumptions when reducing and analysing most HST data, the treatment of the contamination by WASP-77B seems the mostly likely cause of this discrepancy.

Both studies extracted the spectrum using the non-destructive reads rather than the full scan. Extracting WFC3 spatial scan data using the non-destructive reads can be a good way of avoiding contamination from other sources in the field of view. However, given the scan rate used, the separation is not large enough in this case to allow for the spectra to be disentangled in this way. To demonstrate this, Figure \ref{fig:contaminated_extraction} shows example extraction apertures from the second visit using this splitting extraction mode for both forward and reverse scans. WASP-77B clearly contaminates the extraction apertures, and so the resulting light curve contains flux from both stars, demonstrating the need for a corrective factor to be applied.

\begin{figure}
    \centering
    \includegraphics[width=\columnwidth]{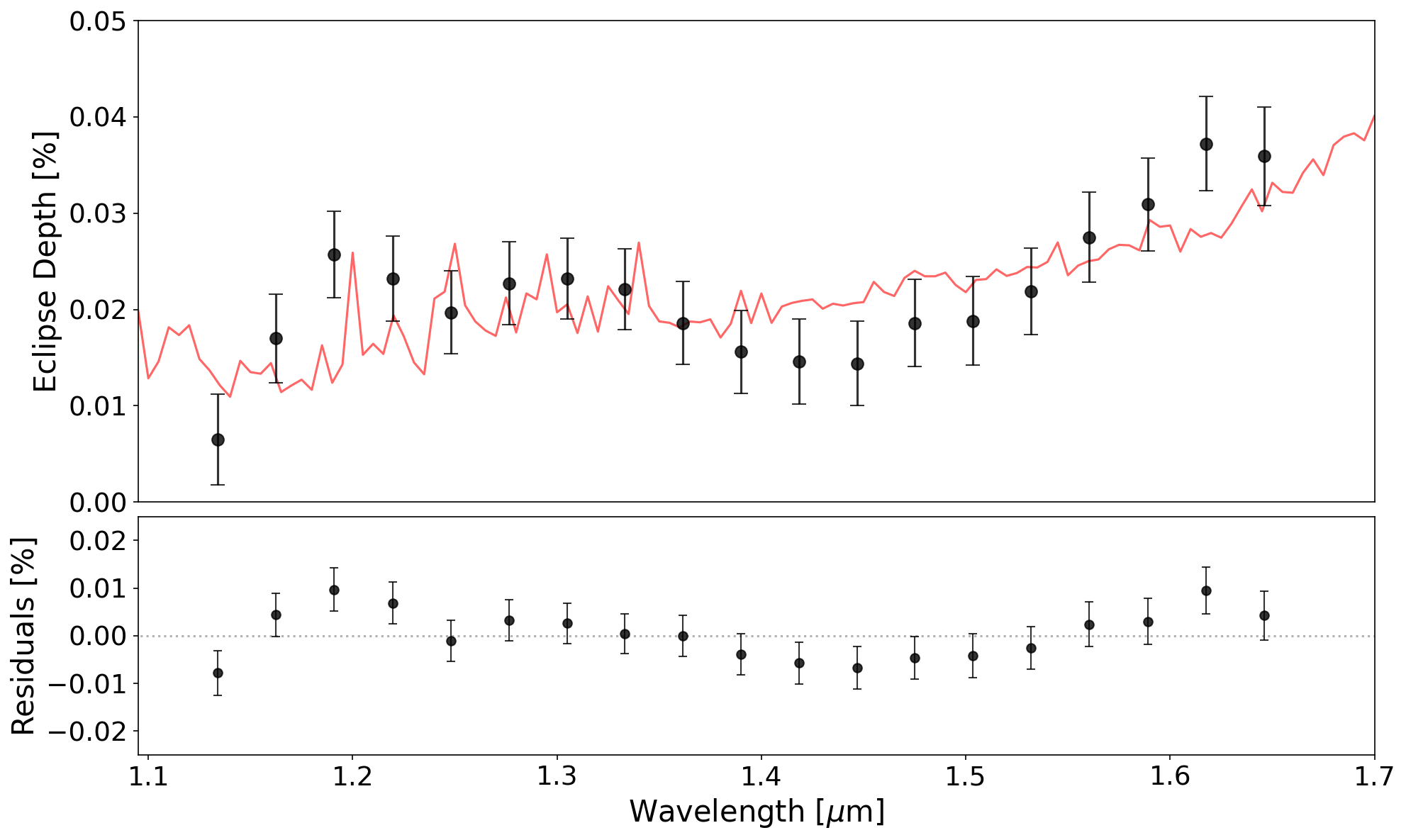}
    \caption{Best-fit model when jointly fitting the JWST NIRSpec data and HST WFC3 data from \citet{mansfield_w77}. The residuals show a clear correlation, indicating a poor fit which is likely due to the contamination present in the spectrum.}
    \label{fig:mansfield_spec}
\end{figure}

In Figure \ref{fig:mansfield_comp}, we show the contaminated eclipse spectrum that is obtained by fitting the light curves of this extraction, as well as the corrected spectrum from \citet{emission_pop} (i.e, after the contamination factor is applied). We note that the same pipeline, Iraclis, is used to extract and fit the light curves in both cases. A clear difference in the two spectra is seen, both in absolute depth and in spectral shape. Both differences are expected: WASP-77B contributes flux to the aperture, thereby reducing the eclipse depth, and this contribution is wavelength dependent.

In the same plot, we also compare the Iraclis spectra to the one reported by \citet{mansfield_w77}, which appears to closely match the contaminated spectrum from Iraclis. The corrected spectrum from \citet{emission_pop}, on the other hand, strongly disagrees with the one from \citet{mansfield_w77}. Looking more closely at methodology utilised by \citet{mansfield_w77}, one finds that they corrected for the contamination by applying a multiplicative factor to the observed flux (see Equation 1). However, this is incorrect as multiplying both the in- and out-of-eclipse data by the same factor does not change the eclipse depth. Instead, one should subtract the flux from the contaminating star.

The analysis of HST WFC3 data is complex. For instance, due to the geometry of the instrument, the spectrum does not scan perfectly vertically, the scan rate changes with position on the detector, and the forward and reverse scans have different lengths \citep[e.g.,][]{mccullough_wfc3_scan,tsiaras_hd209,tsiaras_55cnce}. Hence, removing the flux contribution of a resolvable background star is also difficult and requires specialised software \citep[e.g.,][]{varley_wayne}. It remains, of course, impossible to perfectly correct for such contamination. However, the agreement between findings from the HST WFC3 data of \citet{emission_pop} and the JWST NIRSpec data from \citet{august_w77_jwst}, as well as the ability to jointly fit both datasets, suggest that the methodology employed by \citet{emission_pop}, which accounted for these instrumental complexities, has led to a more trustworthy spectrum. 

To further support this statement, we reproduce the joint retrievals presented in the main text using the HST spectrum from \citet{mansfield_w77}. In this case, the models cannot explain the spectral shape of the WFC3 data, echoing the results of \citet{august_w77_jwst}. Figure \ref{fig:mansfield_spec} shows the best-fit free retrieval for this spectrum and highlights the strong remaining correlations in the residuals. The retrieval also prefers an unphysically large abundance of FeH (i.e, log$_{\rm 10}$(FeH) = -3.26$^{+0.39}_{-0.33}$). The CO and H$_2$O abundances, however, remained consistent with those stated in the main text with log$_{\rm 10}$(CO) = -4.58$^{+0.36}_{-0.29}$, and log$_{\rm 10}$(H$_2$O) = -4.47$^{+0.18}_{-0.10}$. The result suggests, again, that the NIRSpec data drive the robust constraints on CO and H$_2$O, while the abundance of TiO, which is mainly determined from the HST data, should be interpreted with caution.

Overall, our investigations for the HST WFC3 data of WASP-77A\,b reiterate the conclusions from \citet{august_w77_jwst}: the HST spectrum from \citet{mansfield_w77} cannot be reconciled with the JWST NIRSpec data. We suggest that the main reason for this is that their reduction did not properly account for the contamination of the companion star, WASP-77B. 

\section{Table of Results}
\label{sec:appendix_b}

Here we present Table \ref{tab:ratios}, which contains the elemental ratios derived in this work. 


\begin{table}[b]
    \centering
    \caption{Elemental ratios derived from our joint fit to the HST WFC3 and JWST NIRSpec data using free chemistry. Stellar values (i.e., those for WASP-77A) are taken from \citet{polanski_st_abund} while the solar abundances are from \citet{asplund_solar_abund}.}
    \begin{tabular}{cccc}\hline \hline
        Ratio & Value & w.r.t Stellar & w.r.t Solar\\\hline \hline
        C/O & 0.54$^{+0.12}_{-0.12}$ & 0.91-1.43 & 0.76-1.2 \\
        log$_{10}$(Ti/O) & -2.30$^{+0.20}_{-0.23}$ & 18.00-49.58 & 16.05-44.19\\
        log$_{10}$(C/H) & -1.21$^{+0.28}_{-0.26}$ & 0.03-0.12 & 0.03-0.12\\
        log$_{10}$(O/H) & -1.08$^{+0.24}_{-0.17}$ & 0.04-0.10 & 0.04-0.10\\
        log$_{10}$(Ti/H) & 0.38$^{+0.22}_{-0.23}$ & 1.05-2.92 & 1.07-2.99\\
        log$_{10}$(C+O/H) & -1.10$^{+0.28}_{-0.20}$ & 0.04-0.10 & 0.04-0.10 \\
        log$_{10}$(C+O+Ti/H) & -1.10$^{+0.28}_{-0.20}$ & 0.04-0.10 & 0.04-0.10 \\
        \hline \hline
    \end{tabular}
    
    \label{tab:ratios}
\end{table}

\end{document}